2021

# Russian Contribution to Coronary Artery Disease Research: A Scientometric Mapping of Publications

Muneer Ahmad
muneerbangroo@gmail.com

M Sadik Batcha
*Annamalai University*, msbau@rediffmail.com

# Russian Contribution to Coronary Artery Disease Research: A Scientometric Mapping of Publications

Muneer Ahmad[1] Dr. M Sadik Batcha[2]

[1]*Research Scholar, Department of Library and Information Science, Annamalai University, Annamalai nagar, muneerbangroo@gmail.com*
[2]*Research Supervisor & Mentor, Professor and University Librarian, Annamalai University, Annamalai nagar, msbau@rediffmail.com*

**Abstract**

The present study attempts to highlight the research output generated in Russia in coronary artery disease (CAD) research during the period 1990-2019 to understand the distribution of research output, top journals for publications, and most prolific authors, authorship pattern, and citation pattern. This study is based on secondary data extracted from the Science Citation Index (SCI), which is an integral component of the Web of Science. Descriptive and inferential statistical techniques were applied in the study. There were 5058 articles by Russian scholars in coronary artery disease during 1990-2019; they preferred to publish in Russian journals. The research contributions were in the form of research articles, meeting abstracts and reviews with a consistent drop in the number of editorial material and article; proceedings paper with time. Co-authorship was the norm in coronary artery disease research, with a steady increase in the number of multi-author documents in recent years.

## 1. Introduction

Coronary artery disease (CAD) refers to the build-up of atherosclerotic plaque in the blood vessels that supply oxygen and nutrients to the heart (Braunwald & Bonow, 2012). The complex process of atherosclerosis begins early in life and is thought to initiate with dysfunction of endothelial cells that line the coronary arteries; these cells are no longer able to appropriately regulate vascular tone (narrowing or constriction of the vessels) with nitric oxide signaling. Progressive infiltration of the vessel wall by lipoprotein particles carrying cholesterol propagates an inflammatory response by cholesterol-loaded macrophage 'foam cells.' Smooth muscle cells underlying the vessel wall proliferate and lead to the vessel's remodeling that can ultimately lead to a narrowing of the vessel that obstructs blood flow. A myocardial infarction (heart attack) is typically caused when a rupture incites a blood clot in the surface of the plaque; this process deprives the heart muscle downstream of the blood clot of adequate blood flow and leads to cell death (Khera & Kathiresan, 2017).

The prevalence of CAD, also known as coronary heart disease (CHD), has been observed to vary significantly according to geographical locations, ethnicity, and gender (Go et al., 2014). Epidemiological studies on such cardiovascular diseases have provided information that could guide the strategies of prevention and eradication of these diseases both at the individual and population levels (Wong, 2014). Even before the field of cardiovascular epidemiology existed, in Minnesota (United States) the first prospective studies of CAD prevalence in population was conducted in 1946 (Keys et al., 1963). In the seven countries study, the relationships between lifestyle, diet, CAD, and stroke were elucidated (Keys, 1980). This study also indicated that the rates of heart attack and stroke were directly related to the levels of total cholesterol, and this remained constant across different countries and cultures (Epstein, Blackburn, & Gutzwiller, 1996).

## 2. Review of Literature

Numerous studies have been conducted in the areas of Scientometrics, Bibliometrics, and related to it, Webometrics (Ahmad, Batcha, Rashid, & Hafiz, 2018). The discipline has been widely spread through different journals, conference articles, monographs, textbooks, etc., especially in recent decades. Because of the enormous amount of literature available in the field, an attempt has been made to review only significant and contemporary literature on the various aspects of scientometrics research. (Batcha & Ahmad, 2017) obtained the analysis of two journals Indian Journal of Information Sources and Services (IJSS), which is of Indian origin, and Pakistan Journal of Library and Information Science (PJLIS) from Pakistan origin and studied them comparatively with scientometric indicators like year-wise distribution of articles, the pattern of authorship and productivity, degree of collaboration, way of co-authorship, average length of papers, average keywords. The collaboration with foreign authors of both the countries is negligible (1.37% of articles) from India and (4.10% of articles) from Pakistan.

(Ahmad, Batcha, Wani, Khan, & Jahina, 2018) studied Webology journal one of the reputed journals from Iran was explored through scientometric analysis. The study aims to provide a comprehensive analysis regarding the journal like year wise growth of research articles, authorship pattern, author productivity, and subjects taken by the authors over the period of 5 years from 2013 to 2017. The findings indicate that 62 papers were published in the journal during the study period. The articles having collaborative nature were high in number. Regarding the subject concentration of papers of the journal, Social Networking, Web 2.0, Library 2.0, and Scientometrics or Bibliometrics were highly noted.

(Batcha, Jahina, & Ahmad, 2018) has examined the DESIDOC Journal by means of various scientometric indicators like year wise growth of research papers, authorship pattern, subjects, and themes of the articles over five years from 2013 to 2017. The study reveals that 227 articles were published over five years, from 2013 to 2017. The authorship pattern was highly collaborative. The maximum number of items (65 %) have ranged their thought contents between 6 and 10 pages.

(Ahmad & Batcha, 2019) analyzed research productivity in Journal of Documentation (JDoc) for a period of 30 years between 1989 and 2018. Web of Science, a service from Clarivate Analytics, has been consulted to obtain bibliographical data, and it has been analyzed through Bibexcel and Histcite tools to present the datasets. The analysis part deals with local and global citation level impact, highly prolific authors, and their research output, the ranking of prominent institution and countries. In addition to this scientographical mapping of bibliographical data is obtainable through VOSviewer, which is open source mapping software.

(Ahmad & Batcha, in 2019) studied the scholarly communication of Bharathiar University, which is one of the vibrant universities in Tamil Nadu. The study finds out the impact of research produced, year-wise research output, citation impact at local and global level, prominent authors and their total output, top journals of publications, top collaborating countries which collaborate with the university authors, highly industrious departments, and trends in publication of the university during 2009 through 2018. In addition, the study used scientographical mapping of data and presented it through graphs using the VOSviewer software mapping technique.

(Ahmad, Batcha, & Jahina, 2019) quantitatively measured the research productivity in the area of artificial intelligence at the global level over the study period of ten years (2008-2017). The study acknowledged the trends and features of growth and collaboration pattern of artificial intelligence research output. The average growth rate of artificial intelligence per year increases at a rate of 0.862. The multi-authorship pattern in the study is found high, and the average number of authors per paper is 3.31. Collaborative Index is noted to be the highest range in the year 2014 with 3.50. Mean CI during the period of study is 3.24. This is also supported by the mean degree of collaboration at a percentage of 0.83. The mean CC observed is 0.4635. Regarding the application of Lotka's Law of authorship productivity in the artificial intelligence literature, it proved to be a fit for the study.

(Batcha, Dar, & Ahmad, 2019) presented a scientometric analysis of the journal titled "Cognition" for a period of 20 years from 1999 to 2018. The present study was conducted with an aim to provide a summary of research activity in the current journal and characterize its most aspects. The research coverage includes the year-wise distribution of articles, authors, institutions, countries and citation analysis of the journal. The analysis showed that 2870 papers were published in journal of Cognition from 1999 to 2018. The study identified the top 20 prolific authors, institutions, and countries of the journal. Researchers from the USA have made the most percentage of contributions.

(Batcha, Dar, & Ahmad, 2020) conducts a scientometric study of the *Modern Language Journal* literature from 1999 to 2018. A total of 2564 items resulted from the publication name using "Modern Language Journal" as the search term was retrieved from the Web of Science Database. Based on the number of publications during the study period, no consistent growth was observed in the research activities pertaining to the journal. The annual distribution of publications, number of authors, institution productivity, country wise publications and Citations are analyzed. Highly productive authors, institutions, and countries are identified. The results reveal that the maximum number of papers 179 is published in the year 1999. It was also observed that Byrnes H is the most productive, contributed 51 publications and Kramsch C is most cited author in the field, having 543 global citations. The highest number (38.26%) of publications contributed from the USA, and the foremost productive establishment was the University of Iowa.

(Ahmad, Batcha, & Dar, 2020) studied the Brain and Language journal which is an interdisciplinary journal, publishes articles that illustrate the complex relationships among language, brain, and behavior and is one such journal which is concerned with investigating the neural correlates of Language. The study aims at mapping the structure of the *Brain and Language* journal. The journal looks into the intrinsic relationship between language and brain. The study demonstrates and elaborates on the various aspects of the Journal, such as its chronology wise total papers, most productive authors, citations, average citation per paper, institution and country wise distribution of publications for a period of 20 years.

(Ahmad & Batcha, 2020) explores and analyses the trend of world literature on "Coronavirus Disease" in terms of the output of research publications as indexed in the Science Citation Index Expanded (SCI-E) of Web of Science during the period from 2011 to 2020. The study found that 6071 research records have been published on Coronavirus Disease. The various scientometric components of the research records published in the study period were studied. The study reveals the multiple aspects of Coronavirus Disease literature such as year wise distribution, relative growth rate, doubling time of literature, geographical wise, organization wise, language wise, form wise , most prolific authors, and source wise.

(Ahmad & Batcha, 2020) analyzed the application of Lotka's law to the research publication, in the field of Dyslexia disease. The data related to Dyslexia were extracted from web of science database, which is a scientific, citation and indexing service, maintained by Clarivate Analytics. A total of 5182 research publications were published by the researchers, in the field of Dyslexia. The study found out that, the Lotka's inverse square law is not fit for this data. The study also analyzed the authorship pattern, Collaborative Index (CI), Degree of Collaboration (DC), Co-authorship Index (CAI), Collaborative Co-efficient (CC), Modified Collaborative Co-efficient (MCC), Lotka's Exponent value, Kolmogorov-Smirnov Test (K-S Test), Relative Growth Rate and Doubling Time.

(Umar, Ahmad, & Batcha, 2020) studied and focused on the growth and development of Library and Culture research in forms of publications reflected in Web of Science database, during the span of 2010-2019. A total 890 publications were found and the highest 124 (13.93%) publications published in 2019.The analysis maps comprehensively the parameters of total output, growth of production, authorship, institution wise and country-level collaboration patterns, significant contributors (individuals, top publication sources, institutions, and countries).

(Ahmad & Batcha, 2020) studied and examined 4698 Indian Coronary Artery Disease research publications, as indexed in Web of Science database during 1990-2019, with a view to understand their growth rate, global share, citation impact, international collaborative papers, distribution of publications by broad subjects, productivity and citation profile of top organizations and authors, and preferred media of communication.

(Jahina, Batcha, & Ahmad, 2020) study deals a scientometric analysis of 8486 bibliometric publications retrieved from the Web of Science database during the period 2008 to 2017. Data is collected and analyzed using Bibexcel software. The study focuses on various aspect of the quantitative research such as growth of papers (year wise), Collaborative Index (CI), Degree of Collaboration (DC), Co-authorship Index (CAI), Collaborative Co-efficient (CC), Modified Collaborative Co-Efficient (MCC), Lotka's Exponent value, Kolmogorov-Smirnov test (K-S Test).

(Ahmad & Batcha, 2020) analyze Brazil research performance on Coronary Artery Disease as reflected in indexed publications in Web of Science with a view to understand their distribution of research output, top journals for publications, most prolific authors, authorship pattern, and citations pattern on CAD. The results indicate that highest growth rate of publications occurred between the years 1995-1999. University Sao Paulo topped the scene among all institutes. The leading publications were more than ten authored publications. Ramires JAF and Santos RD were found to be the most prolific authors. It is also found that most of the prolific authors (by number of publications) do not emerge in highly cited publications' list. CAD researchers mostly preferred using article publications to communicate their findings.

(Ahmad & Batcha, 2020) presented and attempted to check the applicability of Lotka's Law on South African publication on Coronary artery disease research. The study lights on Lotka's empirical law of scientific productivity, i.e., Inverse Square Law, to measure the scientific productivity of authors, to test Lotka's Exponent value and the K.S test for the fitness of Lotka's Law.

## 3. Objectives

The main objective of the present study is to study the growth of research output on Coronary Artery Disease from Russia. Moreover, the analysis has been performed:

- To find out the type of documents containing Coronary Artery Disease research output in Russia during 1990-2019;
- To analyze the year-wise distribution and growth of literature on Coronary Artery Disease in Russia during 1990-2019;
- To identify the top institutions researching Coronary Artery Disease;
- To determine the most prolific authors exploring Coronary Artery Disease;
- To study the authorship pattern in Coronary Artery Disease research;
- To analyze the top sources preferred by authors for publishing Coronary Artery Disease research.

## 4. Methodology

The present study is a scientometric analysis of Coronary Artery Disease research publications. A total of 5058 records have been extracted from the Web of Science database in the '.txt' format covering the period (1990-2019). The search string used for data extraction is:

"TS=(Artery Disease, Coronary OR Artery Diseases, Coronary OR Coronary Artery Diseases OR Disease, Coronary Artery OR Diseases, Coronary Artery OR Coronary Arteriosclerosis OR Arterioscleroses, Coronary OR Coronary Arterioscleroses OR Atherosclerosis, Coronary OR Atheroscleroses, Coronary OR Coronary Atheroscleroses OR Coronary Atherosclerosis OR Arteriosclerosis, Coronary OR Ischaemic OR Ischemic OR hardening of the Arteries OR Induration of the Arteries OR Arterial Sclerosis ) AND CU=(Russia OR USSR)"

This search has been refined to limit the period from 1990 to 2019. Data filtering has been performed manually to remove irrelevant record entries. Bibliometrix Package in RStudio has been used for analyzing the data and it has also been used for tabulation and visualization of Results.

## 5. Data Analysis and Findings

### 5.1. Type of Publications

Different kind of publications in which research work on Coronary Artery Disease from Russia is contributed during last 30 years is listed in Table 1. Out of total publications 3572 (70.5 %) are research articles, 865 (17.1 %) are meeting abstracts, 466 (9.2 %) are reviews, 64 (1.3 %) are editorial material, 59 (1.2 %) are article; proceedings paper, 12 (0.2 %) are note, 7 (0. 1 %) are letter, 5 (0.1 %) are article; early access, 2 are article; book chapter, Discussion, News Item, and 1 are correction and review & book chapter. It is apparent that more research output was produced in the form of articles, and is having second highest ACPP (16.59) than other forms of publications. It is also evident that despite more research output was produced in articles but ACPP of research output published as article; proceedings paper was also fair amount (28.88) compared to articles (16.59). ACPP of review having (15.29), letter (7.29), note (5.17), and review; book chapter (2.00). Article; Book Chapter publication published on CAD also received 1.50 ACPP. Other type of documents had ACPP less than 1.50. Thus; it was observed that articles, reviews and article; proceedings paper received more citations than other forms of documents.

## Table 1: Publication Type

| S.No. | Document Type | Records | Percentage | TC | ACPP |
|---|---|---|---|---|---|
| 1 | Article | 3572 | 70.5 | 60543 | 16.59 |
| 2 | Meeting Abstract | 865 | 17.1 | 88 | 0.10 |
| 3 | Review | 466 | 9.2 | 7127 | 15.29 |
| 4 | Editorial Material | 64 | 1.3 | 79 | 1.23 |
| 5 | Article; Proceedings Paper | 59 | 1.2 | 1704 | 28.88 |
| 6 | Note | 12 | 0.2 | 62 | 5.17 |
| 7 | Letter | 7 | 0.1 | 51 | 7.29 |
| 8 | Article; Early Access | 5 | 0.1 | 3 | 0.60 |
| 9 | Article; Book Chapter | 2 | 0 | 3 | 1.50 |
| 10 | Discussion | 2 | 0 | 2 | 1.00 |
| 11 | News Item | 2 | 0 | 0 | 0.00 |
| 12 | Correction | 1 | 0 | 0 | 0.00 |
| 13 | Review; Book Chapter | 1 | 0 | 2 | 2.00 |

TC= "Total Citations", ACPP= "Average Citations per Paper"

## 5.2. Distribution of Research Publications

There has been a continuous increase in publications from the first decade (1990-1999) to the latest decade (2010-2019). During 30 years of research, about 54.2 per centre research output on CAD was contributed in decade third (2010-2019). Table 2 shows the distribution of research output in five blocks of five years each. It is very apparent that the highest growth rate occurs in the block year 1995-1999 (154.63%) followed by 2015-2019 (38.62%). Almost one-third (28.97%) research output on CAD was contributed during 1990-2004. In first block year, research output was (4.28%), and in second block it increased (10.90%), and in the third block, it again increased (13.78%) and afterward increased continuously by every block year. The highest number of research was contributed in the partnership 2015-2019 (31.38%).

## Table 2: Distribution of Papers during 1990-2019

| Year | Publications | % of TP | CO | % of Growth |
|---|---|---|---|---|
| 1990-1994 | 216 | 4.28 | 216 | -- |
| 1995-1999 | 550 | 10.90 | 766 | 154.63 |
| 2000-2004 | 695 | 13.78 | 1461 | 26.36 |
| 2005-2009 | 872 | 17.29 | 2333 | 25.47 |
| 2010-2014 | 1142 | 22.64 | 3475 | 30.96 |
| 2015-2019 | 1583 | 31.38 | 5058 | 38.62 |
| Total | 5058 | 100.28 | | |

TP= "Total Publications", CO= "Cumulative Output", Formula of Growth= "Final Value-Start Value/Start Value X100"

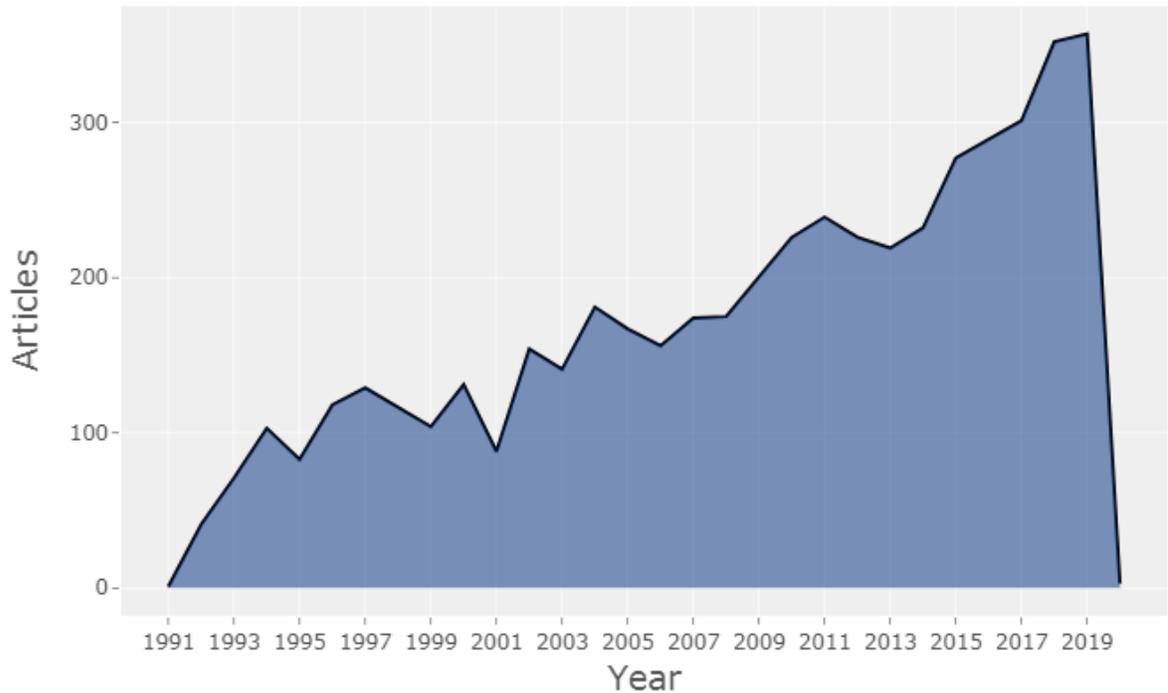

### 5.3. Institution-wise Research Share

The top 20 institutions that produced the highest research outputs on CAD during the period under study are listed in Table 3. Table 3 summarizes total articles, the complete citation score, and average citation per paper of the publications of these institutions. In total, 9294 institutions, including 14925 subdivisions, published 5058 research papers during 1990 – 2019. The top twenty institutions involved in this research have published 53 and more research articles. The mean average is 1.84 research articles per Institution. Out of 9294 institutions, the top 20 institutions published 2424 collaborative research papers. It is also observed that among twenty top Institutions which contributed the highest research output on CAD, Russian Academy Medical Science took the lead by producing a research output of 576 publications followed by Russian Academy Science with 417 research publications followed by Russian State Medical University with 228 research publications followed by Moscow MV Lomonosov State University with 163 research publications. Sixteen institutions produced 50 or more than 500 research publications on CAD. In terms of citations, Russian Academy Science received the highest citations i.e. 9675 for 417 total research publications. It is also noticed that Minist Hlth Russian Federat, has highest ACPP (36.67).

Table 3: Top Institutions Research Output

| S.No. | Institution | Records | Percent | TC | ACPP |
|---|---|---|---|---|---|
| 1 | Russian Academy Medical Science | 576 | 11.4 | 5799 | 10.07 |
| 2 | Russian Academy Science | 417 | 8.3 | 9675 | 23.20 |
| 3 | Russian State Medical University | 228 | 4.5 | 2112 | 9.26 |
| 4 | Moscow MV Lomonosov State University | 163 | 3.2 | 4990 | 30.61 |
| 5 | Minist Hlth Russian Federat | 89 | 1.8 | 3264 | 36.67 |
| 6 | IM Sechenov Medical Academy | 84 | 1.7 | 174 | 2.07 |
| 7 | Cardiol Research Centre | 82 | 1.6 | 2604 | 31.76 |
| 8 | National Research Centre Prevent Medical | 78 | 1.5 | 1295 | 16.60 |
| 9 | Russian Cardiol Research & Prod Complex | 68 | 1.3 | 790 | 11.62 |
| 10 | Cardiol Research Complex | 66 | 1.3 | 368 | 5.58 |
| 11 | Pirogov Russian National Research Medical University | 65 | 1.3 | 503 | 7.74 |
| 12 | IM Sechenov First Moscow State Medical University | 64 | 1.3 | 297 | 4.64 |
| 13 | Ministry Public Health Russia | 59 | 1.2 | 309 | 5.24 |
| 14 | AL Myasnikov Clin Cardiol Institution | 58 | 1.1 | 80 | 1.38 |
| 15 | RAMS | 58 | 1.1 | 162 | 2.79 |
| 16 | AL Myasnikov Cardiol Institution | 56 | 1.1 | 1255 | 22.41 |
| 17 | Russian Cardiol Science & Production Centre | 54 | 1.1 | 200 | 3.70 |
| 18 | Research Centre Prevent Medical | 53 | 1.1 | 541 | 10.21 |
| 19 | Research Institution Complex Issues Cardiovascular Disease | 53 | 1.1 | 98 | 1.85 |
| 20 | Siberian State Medical University | 53 | 1.1 | 93 | 1.75 |

TP= "Total Publications", TC= "Total Citations", ACPP= "Average Citations per Paper".

## 5.4. Most Prolific Authors

The list of twenty top authors who produced the highest contribution to research output on CAD in Russia is given in Table 4. In terms of the number of publications, Barbarash OL is the most productive author with 100 publications, followed by Skvortsova VI 93, Orekhov AN 84, and Belenkov YN 83 publications. It is also noted that 1 out of 20 prolific authors contributed more than a hundred research publications each while 19 authors contributed more than 38 journals each. The h index is highest for Orekhov AN (22), followed by Sobenin IA (17), followed by Skvortsova VI (11) and Pokushalov E, & Romanov A (10). The data set puts forth the authors Orekhov AN with 33 g-index, Sobenin IA with 25 g-index, Pokushalov E with 24 g-index, Barbarash OL, and Ezhov MV with 23 g-index, Belenkov YN and Romanov A with 20 g-index, and Skvortsova VI with 15 g-index. Romanov A (0.7692), Orekhov AN (0.7586), Sobenin IA (0.5862) are having the highest m- index, respectively.

Table 4: Most Prolific Authors

| S.No. | Author | TP | TC | h-index | g-index | m-index |
|---|---|---|---|---|---|---|
| 1 | Barbarash OL | 100 | 620 | 8 | 23 | 0.3200 |
| 2 | Skvortsova VI | 93 | 403 | 11 | 15 | 0.4231 |
| 3 | Orekhov AN | 84 | 1396 | 22 | 33 | 0.7586 |
| 4 | Belenkov YN | 83 | 494 | 9 | 20 | 0.3214 |
| 5 | Sidorenko BA | 78 | 121 | 5 | 8 | 0.1852 |
| 6 | Kukharchuk VV | 63 | 208 | 8 | 12 | 0.2759 |
| 7 | Gratsiansky NA | 55 | 120 | 5 | 7 | 0.1724 |
| 8 | Deev AD | 52 | 145 | 7 | 10 | 0.2500 |
| 9 | Masenko VP | 49 | 120 | 5 | 7 | 0.2174 |
| 10 | Pokushalov E | 46 | 583 | 10 | 24 | 0.3200 |
| 11 | Karpov YA | 46 | 80 | 4 | 7 | 0.1429 |
| 12 | Ezhov MV | 44 | 553 | 9 | 23 | 0.3600 |
| 13 | Kuznetsov VA | 43 | 31 | 3 | 3 | 0.1304 |
| 14 | Sobenin IA | 43 | 710 | 17 | 25 | 0.5862 |
| 15 | Lyakishev AA | 42 | 119 | 5 | 9 | 0.1724 |
| 16 | Oganov RG | 42 | 170 | 8 | 11 | 0.2857 |
| 17 | Romanov A | 42 | 423 | 10 | 20 | 0.7692 |
| 18 | Akchurin RS | 41 | 129 | 5 | 10 | 0.1852 |
| 19 | Karpov RS | 39 | 112 | 6 | 9 | 0.2222 |
| 20 | Gusev EI | 38 | 201 | 8 | 13 | 0.3077 |

TP= "Total Publications", TC= "Total Citations", ACPP= "Average Citations per Paper".

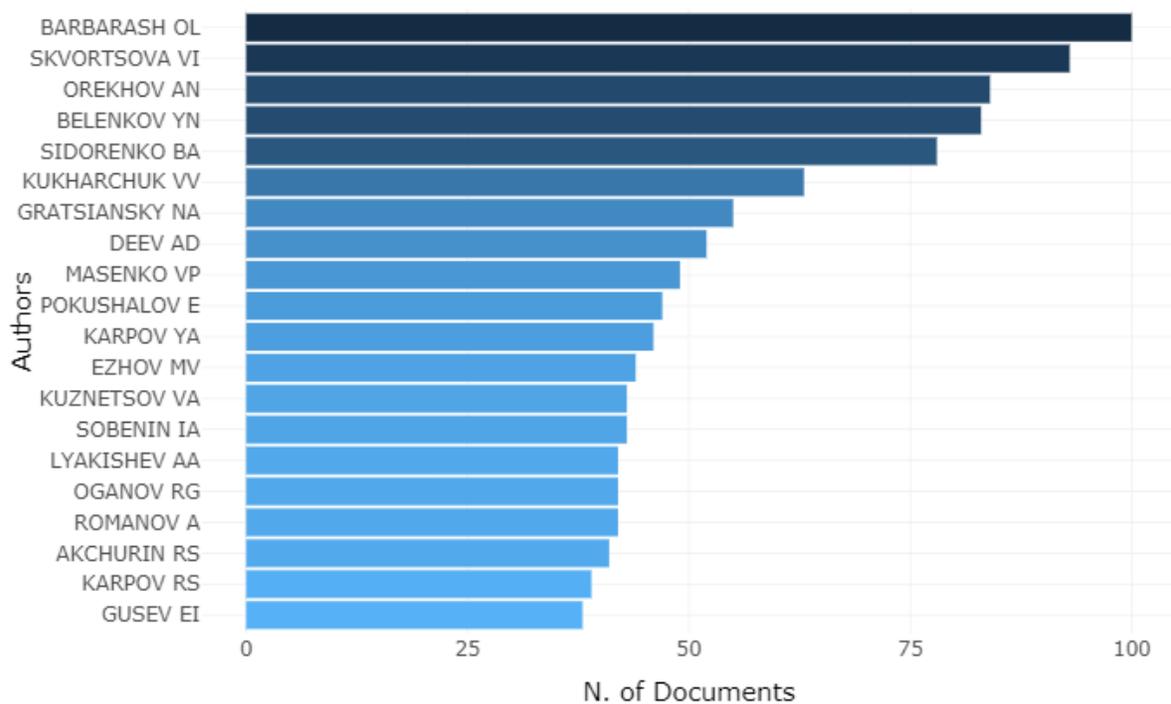

## 5.5. Authorship Pattern

Table 5 illustrates the overall and five year-wise distribution of authorship trends. It is evident from Table 5 that only 6.70 percent of publications were single-authored publications, while the rest of 93.30 had two or more authors. The maximum number of publications were four authored publications (14.69%), nearly followed by five written publications (13.88%), three authored (13.09%), six authored (11.66%), and seven authored publications (8.68%). Two to nine written publications accounted for 82.44 percent, while more than ten authored publications accounted for 8.05 percent.

Table 5: Authorship Pattern

| Author(s) | Total Research Output (5 Yearly) | | | | | | Total Research Output | |
|---|---|---|---|---|---|---|---|---|
| | 1990-1994 | 1995-1999 | 2000-2004 | 2005-2009 | 2010-2014 | 2015-2019 | Total | % |
| Single | 22 | 51 | 62 | 69 | 86 | 49 | 339 | 6.70 |
| Two | 34 | 70 | 87 | 98 | 122 | 127 | 538 | 10.64 |
| Three | 36 | 91 | 110 | 122 | 153 | 150 | 662 | 13.09 |
| Four | 35 | 97 | 102 | 122 | 159 | 228 | 743 | 14.69 |
| Five | 33 | 70 | 98 | 125 | 142 | 234 | 702 | 13.88 |
| Six | 32 | 62 | 86 | 99 | 134 | 177 | 590 | 11.66 |
| Seven | 12 | 42 | 60 | 86 | 110 | 129 | 439 | 8.68 |
| Eight | 6 | 30 | 25 | 55 | 75 | 105 | 296 | 5.85 |
| Nine | 0 | 11 | 27 | 31 | 50 | 81 | 200 | 3.95 |
| Ten | 2 | 10 | 15 | 22 | 28 | 65 | 142 | 2.81 |
| More than 10 | 4 | 16 | 23 | 43 | 83 | 238 | 407 | 8.05 |
| Total | 216 | 550 | 695 | 872 | 1142 | 1583 | 5058 | 100.00 |
| % | 4.27 | 10.87 | 13.74 | 17.24 | 22.58 | 31.30 | | |

## 5.6. Top Journals Preferred for Publication

The total number of 5058 publications on CAD from 1990 to 2019 appeared in 628 different sources. The top 20 journals preferred for CAD publications are listed in Table 6, which accounted for 26.89 percent of total research publications during the period under study. Kardiologiya has published the highest (1360) publications on CAD, followed by Terapevticheskii Arkhiv (543). According to the journals preferred for publication output from the table 6 the journal wise distribution of research documents, Kardiologiya has the highest number of research documents 1360 with 1970 of total citation score and 13, 18 and 0 .448 h index, g index and m index respectively and being prominent among the 20 journals and it stood in first rank position. Terapevticheskii Arkhiv has 543 research documents, and it stood in the second position with 466 of total citation score, and 6, 9, 0.207 h index, g index, and m index score were scaled. It is followed by Bulletin of Experimental Biology and Medicine with 250 records, and it stood in third rank position along with 585 of total citation score and 10, 14, and 0.345 h, g, and m index score measured.

**Table 6: Top 20 Sources for Publications**

| S.No. | Source of Publication | NP | TC | h-index | g-index | m-index |
|---|---|---|---|---|---|---|
| 1 | Kardiologiya | 1360 | 1970 | 13 | 18 | 0.448 |
| 2 | Terapevticheskii Arkhiv | 543 | 466 | 6 | 9 | 0.207 |
| 3 | Bulletin of Experimental Biology and Medicine | 250 | 585 | 10 | 14 | 0.345 |
| 4 | Zhurnal Nevrologii I Psikhiatrii Imeni S S Korsakova | 239 | 312 | 7 | 9 | 0.438 |
| 5 | Cardiovascular Therapy and Prevention | 153 | 84 | 4 | 5 | 0.286 |
| 6 | European Heart Journal | 133 | 2083 | 10 | 45 | 0.385 |
| 7 | Zhurnal Nevropatologii I Psikhiatrii Imeni S S Korsakova | 131 | 295 | 9 | 11 | 0.321 |
| 8 | European Journal of Heart Failure | 109 | 222 | 3 | 14 | 0.214 |
| 9 | Atherosclerosis | 108 | 226 | 9 | 14 | 0.333 |
| 10 | Russian Journal of Cardiology | 64 | 45 | 4 | 4 | 0.286 |
| 11 | Journal of Hypertension | 61 | 242 | 5 | 15 | 0.238 |
| 12 | European Journal of Neurology | 48 | 196 | 4 | 14 | 0.174 |
| 13 | Atherosclerosis Supplements | 47 | 104 | 3 | 10 | 0.167 |
| 14 | Russian Journal of Genetics | 37 | 44 | 3 | 4 | 0.143 |
| 15 | Cerebrovascular Diseases | 35 | 78 | 3 | 8 | 0.111 |
| 16 | Journal of The American College of Cardiology | 34 | 974 | 5 | 31 | 0.313 |
| 17 | Biochemistry-Moscow | 33 | 624 | 13 | 24 | 0.448 |
| 18 | Circulation | 33 | 1272 | 10 | 33 | 0.333 |
| 19 | European Journal of Nuclear Medicine And Molecular Imaging | 32 | 1002 | 9 | 30 | 0.231 |
| 20 | Stroke | 31 | 1019 | 14 | 31 | 0.609 |

NP= "Number of Publications", TC= "Total Citations"

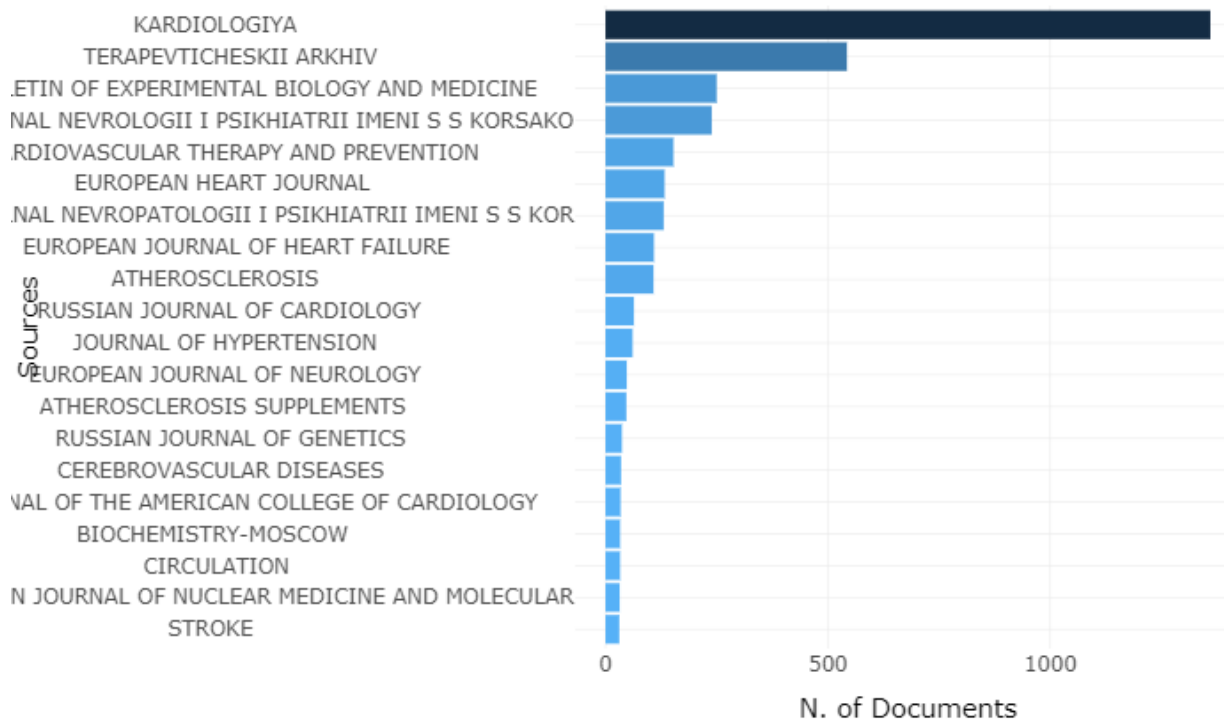

## 6. Conclusion

The study explores the 30 years of research output on CAD in Russian. It was found that a total number of 5058 papers on CAD were published during 1990-2019, which received 69664 citations with an ACPP of 13.77. The growth rate was highest (154.63%) in the block year 1995-1999. ACPP of Minist Hlth Russian Federat has the highest with 36.67 average citations per paper. Nearly 26.89 percent of research on CAD was published in 20 journals, among which Kardiologiya produced the highest research output on CAD. Barbarash OL and Skvortsova VI were the front runners in several publications, but citations and h-index Orekhov AN remained at the top. Only 6.70 percent of publications were single-authored publications, while the rest of 93.30 has two or more authors. Among all types of publications, articles and reviews received more citations. The study depicts that research on CAD in Russia was significantly less in earlier years or decades but increased during the later decades. Significant research output was produced near the 21st century, especially during the last decade.

# References

Ahmad, M., & Batcha, M. S. (2019). Mapping of Publications Productivity on Journal of Documentation 1989-2018 : A Study Based on Clarivate Analytics – Web of Science Database. *Library Philosophy and Practice (e-Journal)*, 2213–2226.

Ahmad, M., & Batcha, M. S. (2019). Scholarly Communications of Bharathiar University on Web of Science in Global Perspective: A Scientometric Assessment. *Research Journal of Library and Information Science*, *3*(3), 22–29.

Ahmad, M., & Batcha, M. S. (2020). Coronary Artery Disease Research in India: A Scientometric Assessment of Publication during 1990-2019. *Library Philosophy and Practice (e-Journal)*. Retrieved from https://digitalcommons.unl.edu/libphilprac/4178

Ahmad, M., & Batcha, M. S. (2020). Examining the Scientific Productivity of Authors in Dyslexia Research: A Study Using Lotka's Law. *Library Philosophy and Practice (e-Journal)*. Retrieved from https://digitalcommons.unl.edu/libphilprac/4198

Ahmad, M., & Batcha, M. S. (2020). Identifying and Mapping the Global Research Output on Coronavirus Disease: A Scientometric Study. *Library Philosophy and Practice (e-Journal)*. Retrieved from https://digitalcommons.unl.edu/libphilprac/4125

Ahmad, M & Batcha, M. S. (2020). Lotka's Law and Authorship Distribution in Coronary Artery Disease Research in South Africa. *Library Philosophy and Practice (e-journal)*. 4457. https://digitalcommons.unl.edu/libphilprac/4457

Ahmad, M & Batcha, M. S. (2020). Measuring Research Productivity and Performance of Medical Scientists on Coronary Artery Disease in Brazil: A Metric Study. *Library Philosophy and Practice (e-journal)*. 4358. Retrieved from https://digitalcommons.unl.edu/libphilprac/4358

Ahmad, M., Batcha, M. S., & Dar, Y. R. (2020). Research Productivity of the Brain and Language Journal from 1999 -2018 : A Scientometric Analysis. *Journal of Advancements in Library Sciences*, *7*(1), 91–99.

Ahmad, M., Batcha, M. S., & Jahina, S. R. (2019). *Testing Lotka' s Law and Pattern of Author*

*Productivity in the Scholarly Publications of Artificial Intelligence*. *Library Philosophy and Practice (e-Journal)* 2716. Retrieved from https://digitalcommons.unl.edu/libphilprac/2716

Ahmad, M., Batcha, M. S., Rashid, W., & Hafiz, O. (2018). Calculating Web Impact Factor for University Websites of Jammu and Kashmir: A Study. *International Journal of Science, Technology and Management (IJSTM)*, *7*(05), 17–27.

Ahmad, M., Batcha, M. S., Wani, B. A., Khan, M. I., & Jahina, S. R. (2018). Research Output of Webology Journal ( 2013-2017 ): A Scientometric Analysis. *International Journal of Movement Education and Social Science*, *7*(3), 46–58.

Batcha, M. S., & Ahmad, M. (2017). Publication Trend in an Indian Journal and a Pakistan Journal: A Comparative Analysis using Scientometric Approach. *Journal of Advances in Library and Information Science*, *6*(4), 442–449.

Batcha, M. S., Dar, Y. R., & Ahmad, M. (2019). Impact and Relevance of Cognition Journal in the Field of Cognitive Science: An Evaluation. *Research Journal of Library and Information Science*, *3*(4), 21–28.

Batcha, M. S., Dar, Y. R., & Ahmad, M. (2020). Global Research Trends in the Modern Language Journal from 1999 to 2018: A Data-Driven Analysis. *Research Journal of Library and Information Science*, *4*(2), 1–8.

Batcha, M. S., Jahina, S. R., & Ahmad, M. (2018). Publication Trend in DESIDOC Journal of Library and Information Technology during 2013-2017: A Scientometric Approach. *International Journal of Research in Engineering, IT and Social Sciences*, *8*(04), 76–82.

Braunwald, E., & Bonow, R. O. (2012). *Braunwald's heart disease : a textbook of cardiovascular medicine*. Philadelphia: Saunders.

Egghe, L. (2006). Theory and practise of the g-index. *Scientometrics*, *69*(1), 131–152. https://doi.org/10.1007/s11192-006-0144-7

Epstein, F. H., Blackburn, H., & Gutzwiller, F. (1996). Cardiovascular disease epidemiology: A journey from the past into the future. *Circulation*, *93*(9), 1755–1764. https://doi.org/10.1161/01.CIR.93.9.1755

Go, A. S., Mozaffarian, D., Roger, V. L., Benjamin, E. J., Berry, J. D., Blaha, M. J., Turner, M. B. (2014). Executive summary: heart disease and stroke statistics--2014 update: a report from the American Heart Association. *Circulation*, *129*(3), 399–410. https://doi.org/10.1161/01.cir.0000442015.53336.12

Hirsch, J. E. (2010). An index to quantify an individual's scientific research output that takes into account the effect of multiple coauthorship. *Scientometrics*, *85*(3), 741–754. https://doi.org/10.1007/s11192-010-0193-9

Jahina, S. R., Batcha, M. S., & Ahmad, M. (2020). Lotka's Law and Pattern of Author Productivity in the Field of Brain Concussion Research: A Scientometric Analysis. *Library Philosophy and Practice (e-Journal)*. Retrieved from https://digitalcommons.unl.edu/libphilprac/4126

Keys, A. (1980). *Seven countries. A multivariate analysis of death and coronary heart disease.* London: Harvard University Press.

Keys, A., Taylor, H. L., Blackburn, H., Brozek, J., Anderson, J. T., & Simson, E. (1963). Coronary Heart Disease Among Minnesota Business and Professional Men Followed Fifteen Years. *Circulation*, *28*(3), 381–395. https://doi.org/10.1161/01.CIR.28.3.381

Khera, A. V, & Kathiresan, S. (2017). Genetics of coronary artery disease: discovery, biology and clinical translation. *Nature Reviews. Genetics*, *18*(6), 331–344. https://doi.org/10.1038/nrg.2016.160

Umar, A. A., Ahmad, M., & Batcha, M. S. (2020). Library and Culture: A Scientometric Analysis and Visualization of Research Trends. *Journal of Cultural and Social Anthropology*, *2*(2), 1–08.

Wong, N. D. (2014). Epidemiological studies of CHD and the evolution of preventive cardiology. *Nature Reviews. Cardiology*, *11*(5), 276–289. https://doi.org/10.1038/nrcardio.2014.26